\documentstyle[prl,aps,epsf,twocolumn,floats]{revtex}
\tighten

\renewcommand{\refname}{}
\newcommand{\biblabel}[1]{[#1]} %
\renewcommand{\references}{%
\ifpreprintsty
\vspace*{-0.1 truein}
\hbox to\hsize{\hss\large \refname\hss}%
\else
\vskip3pt
\hrule width\hsize\relax
\vskip -0.2in
\fi
\list{\biblabel{\arabic{enumiv}}}%
{\labelwidth\WidestRefLabelThusFar  \labelsep4pt %
\leftmargin\labelwidth %
\advance\leftmargin\labelsep %
\ifdim\baselinestretch pt>1 pt %
\parsep  4pt\relax %
\else %
\parsep  0pt\relax %
\fi
\itemsep\parsep %
\usecounter{enumiv}%
\def\theenumiv{\arabic{enumiv}}%
}%
\let\newblock\relax %
\sloppy\clubpenalty4000\widowpenalty4000
\sfcode`\.=1000\relax
\ifpreprintsty\else\small\fi
}

\begin{document}

\draft

\title{
Sweeping of Lattice Disorder and Associated Phenomena in Colossal
Magnetoresistance Compounds.
}

\author{Evgenii E. Narimanov$^1$, and Chandra M. Varma$^2$}

\address{${}^1$ Electrical Engineering Department, Princeton University,
Princeton NJ 08540, \\ ${}^2$ Bell Laboratories-- Lucent Technologies,
700 Mountain Ave., Murray Hill NJ 07974}

\date{ \today}

\maketitle

\begin{abstract}
We show through  a variational calculation that in a large range of parameters
the paramagnetic to
ferromagnetic  transition in colossal magnetoresistance
compounds is accompanied
by a collapse of polaronic lattice disorder in addition to that of spin disorder.
The spin-lattice disordered state is shown to be localized and the ordered state
itinerant as observed.
The observed dramatic change
in the diffuse scattering at the transition as well as the isotope effect
are also explained.
\end{abstract}

\vspace*{-0.05 truein}


\vspace*{-0.15 truein}

While the basic physics of the colossal magnetoresistance (CMR)  phenomena\cite{first_cmr,Ramirez}
follows \cite{first_de_in_cmr,Furukawa98,ShengPRL,ShengPRB} 
 from the Double-Exchange (DE)
model, \cite{Zener,Anderson,deGennes}, several anomalous results have
emerged in a number of recent experiments\cite{Shimomura99,Vasiliu99,Zhao96},
which require additional considerations:
(i) the high-temperature paramagnetic phase in
these compounds is often characterized by insulating behavior;
(ii) the ferromagnetic phase transition observed in these materials
is often first order; (iii) the paramagnetic phase is characterized by an
intense diffuse scattering,
which disappears with temperature in a manner
correlated with the spontaneous magnetization; and (iv) the transition temperature displays
huge isotope effects. These phenomena are related in that when there is
little diffuse scattering,
the transition is second order or weakly first order, 
the resistivity above the transition is
metallic and there is only a small isotope effect.

The diffuse scattering experiments reveal inhomogeneous lattice configurations
in the paramagnetic phase which anneal dramatically in the vicinity of the
CMR transition\cite{Shimomura99,Vasiliu99}. Spin and lattice ``disorder" in
CMR compounds are thus closely related. We believe this provides
a important clue to the essential physics of the colossal magnetoresistance. In the
present Letter, we present a systematic method to describe this effect and its consequences.

The problem we solve also has a more general context. There are two
different ways in which electron-lattice interactions are considered.
One, which we will call the polaron or the Holstein approach,\cite{polarons}
first deals with the problem of a
localized electron interacting with the lattice to form a polaron,
then the hopping from site to site is introduced, and the polaron
Bloch-states are formed.\cite{Martin,YuMin99} In the other approach, which may be associated with
Migdal,\cite{Migdal}
one first
forms the Bloch-states of the electrons, and then treats the electron-phonon
interactions perturbatively in $\Theta/E_F$, where $\Theta$ is the Debye
temperature and $E_F$ is the Fermi energy. These two methods yield vastly
different results, the electron self-energies being typically of the
order of the Fermi energy in the first case, and $O\left(\Theta\right)$
in the second. The transition behavior between these two limits is however not
yet adequately understood, and has recently attracted considerable 
attention.\cite{millis96}

The model we consider is defined by the Hamiltonian
\begin{eqnarray}
H & = & \sum_{\langle ij\rangle} t_{ij}^0
\cos\frac{\theta_{ij}}{2} c^\dagger_i c_j +
\sum_i \left[ 
- \lambda u_i c^\dagger_i c_i + \frac{\hat{p}_i^2}{2 m} + \frac{k}{2}  u_i^2  \right]
\label{eq:H}
\end{eqnarray}
The first three term represent the standard double-exchange (DE)
model with
transfer integrals $t_{ij}^{\rm eff} \equiv t_{ij}^0 \cos\frac{\theta_{ij}}{2}$ which depend on $\theta_{ij}$,
the mutual orientation of spins at sites
$i$ and $j$. The remaining terms in Eq. (\ref{eq:H})
describe the electron-lattice interaction, and a model  lattice
Hamiltonian. Due to the large difference between the ionic radius of ${\rm Mn}^{3+}$ and
${\rm Mn}^{4+}$, the breathing mode of the  $O_6$ octahedra is expected to be the most
important, although asymmetric or Jahn-Teller modes may also
contribute.\cite{millis95} The dynamics of the oxygen octahedra surrounding manganese ions, is
represented in our model by localized classical
oscillators with a scalar displacement $u_i$ from some reference point and
the corresponding momentum $p_i$.
The principal physics underlying the CMR phenomenon, can be uncovered using
one coordinate per each ${\rm Mn}$ ion. Similar models for the
CMR problem have been considered before.\cite{millis96}

The essential physical point we focus on is the competition
between the electrostatic energy,
which favors large lattice distortions provided the electrons are localized,
and the gain in kinetic energy due to itinerancy of electrons which suppresses lattice
distortions. This competition is self-consistently reinforced by
spin-disorder favored by entropy and spin-order favored by the kinetic energy.
(Extrinsic disorder tilts this competition in favor of localized states.)
Accordingly we consider a variational density matrix:
\begin{eqnarray}
\rho & \propto & \exp\left( - \frac{1}{T} \sum_k \left(\varepsilon_k - \mu\right) c_k^\dagger c_k \right) \Pi_j \rho_j
\label{eq:rho}
\end{eqnarray}
where $\mu$ is the chemical potential. The energies $\varepsilon_k$ are  determined variationally, and
\begin{eqnarray}
\rho_j & = &
\sum_{u_1,u_2}  \sum_{\vartheta_1, \vartheta_2} P\left(u_1, u_2; \vartheta_1, \vartheta_2\right)
\left[ c_j^\dagger c_j \left| j, u_1, \vartheta_1 \rangle \langle j, u_1, \vartheta_1 \right| 
\right. \nonumber \\
&  + & \left. \left(1 - c_j^\dagger c_j \right)
\left| j, u_2, \vartheta_2 \rangle \langle j, u_2, \vartheta_2 \right|
\right] \label{eq:P}
\end{eqnarray}
where the operator $\left| j, u, \vartheta \rangle \langle j, u, \vartheta
\right|$ represents a projection
on a state with the lattice distortion $u$ and spin direction $\vartheta$
at site $j$, while $u_1$ and $u_2$ are the lattice distortions
at ``occupied" and ``empty" sites respectively. The probability distribution
 P is also determined variationally as described below;
different limits of it describe paramagnetic localized electronic states with
 distorted lattice configurations or itinerant
ferromagnetic states with uniform lattice configurations.

Neglecting local spin-lattice correlations
we represent $P$ as :
\begin{eqnarray}
P\left(u_1, u_2; \vartheta_1, \vartheta_2\right) & = & P_u\left(u_1,u_2\right)
P_\vartheta\left(\vartheta_1\right) P_\vartheta\left(\vartheta_2\right)
\label{eq:p_product}
\end{eqnarray}
Substituting (\ref{eq:p_product}) into the free energy obtained from
Eq. (\ref{eq:rho}), and taking the functional
derivative with respect to the lattice distortion distribution $P_u$,
after a straightforward but lengthy calculation
we find  that 
 the energies $\varepsilon_k$
correspond to the eigenstates of the Hamiltonian
\begin{eqnarray}
H_e & = & \langle \cos\frac{\theta}{2} \rangle 
\langle \exp\left( - \frac{k \delta u^2}{\hbar \omega_0} \right) \rangle
\sum_{ij} t_{ij}^0   c_i^\dagger c_j \rightarrow \sum_k \varepsilon_k c_k^\dagger c_k
\label{eq:h_e}  
\end{eqnarray}
where $\delta u \equiv u_2 - u_1$.

The variational procedure yields the physically transparent result for the lattice
distortions distribution  
\begin{eqnarray}
P_u\left(u_1, u_2\right) & \propto & \exp\left( - E\left[u_1, u_2\right]/T \right)
\end{eqnarray}
so that the effective free energy
\begin{eqnarray}
F\left[P_\vartheta\right] & = & -
T \log\left[ \int du_1 \int du_2 \exp
\left( - \frac{E\left[u_1, u_2\right]}{T} \right) \right]
\nonumber \\
& + & T \int d\vartheta \sin\vartheta
P_\vartheta\left(\vartheta\right) \log P_\vartheta\left(\vartheta\right)
\label{eq:Fm}
\end{eqnarray}
where
\begin{eqnarray}
E\left[ u_1, u_2 \right] & = & \int_{-\infty}^\mu d\varepsilon \rho\left(\varepsilon\right) \varepsilon
- \lambda u_1 x + E_L\left(u_1, u_2\right)
\label{eq:e_u}
\end{eqnarray}
and the lattice energy
\begin{eqnarray}
E_L\left(u_1, u_2\right) & = &
\frac{k}{2}\left(x u_1^2 + \left(1 - x\right) u_2^2\right).
\label{eq:E_L}
\end{eqnarray}
 $x$ is the number of electrons per site,
$\rho\left(\varepsilon\right)$ is the electronic density of states (DOS). The first
term in Eq. (\ref{eq:e_u}),
\begin{eqnarray}
 \int_{-\infty}^\mu d\varepsilon \rho\left(\varepsilon\right) \varepsilon
\simeq  - \frac{c_0}{2} x \left(1 - x\right) W, 
\nonumber
\end{eqnarray}
where $c_0$ is a constant which depends on the functional form of the density of states
(e.g. for
rectangular DOS $c_0 = 1$, while for the Gaussian functional form of
$\rho\left(\varepsilon\right)$
we find $c_0 \approx 2.26$ and for "elliptical" DOS
$c_0 \approx 1.7$).
The
bandwidth $W$ depends both on the spin distribution and the lattice distortions:
\begin{eqnarray}
W \simeq W_\theta  \exp\left( - \frac{k \ \delta u^2}{\hbar\omega_0} \right)
\label{eq:W}
\end{eqnarray}
where $W_\theta \equiv W_0 \langle \cos\frac{\theta}{2} \rangle$.
For large $\delta u$ the Eqn. (\ref{eq:W}) represents the suppression of
 the bandwidth
due to small polaron  formation.

The critical temperature $T_c$ for the CMR transition is typically
much smaller than the electronic cohesive energy.
Therefore, to adequately describe phenomena in the vicinity of the transition and below
it
we  need consider only the limit $T \ll E\left(u_1, u_2\right)$. Then the
integral over the distortions $u_1$ and $u_2$ in the effective
 free energy Eq. (\ref{eq:Fm}) can
be calculated by the method of steepest descent.
Changing the variables to $\overline{u} \equiv \frac{1}{2}
\left(u_1 + u_2\right)$ and $\delta u$, we find
\begin{eqnarray}
F & = & - T \log \left[
\sum_j \left| \frac{\partial^2 E_u}{\partial \delta u^2} \right|^{-1/2}
\exp\left( - \frac{E_u\left(\delta u^{(j)}\right)}{T} \right)
\right] 
\nonumber \\
& + & T \int d\vartheta \sin\vartheta P_\vartheta\left(\vartheta\right) \log P_\vartheta\left(\vartheta\right)
\label{eq:F_du}
\end{eqnarray}
where $E_u$ depends only on $\delta u$ :
\begin{eqnarray}
E_u\left(\delta u\right) & = & x \left(1 - x\right) \left[- \frac{c_0}{2} W\left(\delta u\right)  -  
\lambda \delta u +  \frac{k}{2} \delta u^2 \right]
\label{eq:E_du}
\end{eqnarray}
and the index $j$ labels different minima of the effective energy $E_u$.

\begin{figure}[t]
\begin{center}
\leavevmode
\epsfxsize = 7. truecm
\epsfbox{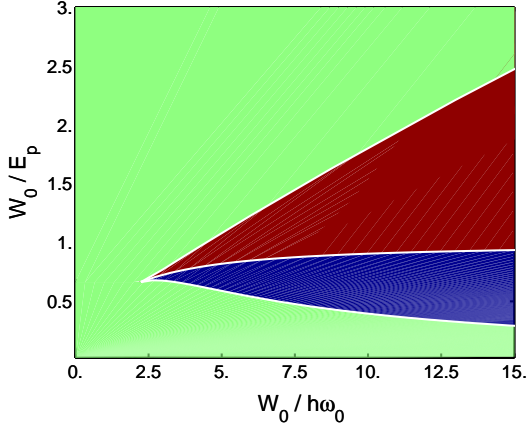}
\end{center}
\caption{
The ``phase diagram" corresponding to the energy 
$E_u\protect\left(\delta u\protect\right)$, which represents the number of energy minima as a function of $W_0/\hbar \omega_0$ and
$W_0 / \protect\left(\lambda^2 / k\protect\right)$.
The green area corresponds to a singe minimum of $E_u$, while the blue and the red 
regions represent the parameters when the energy has two minima with the lowest being
either the Holstein minimum (the blue region) or Migdal minimum (the red region).
\protect\label{fig:phase_diagram}
}
\end{figure}

Depending on the ratios of the electron bandwidth $W_0$ to the polaron energy
$E_p \equiv \lambda^2/k$ and the phonon energy $\hbar \omega_0$,   the
function $E_u\left(\delta u\right)$ has either one or two minima. The corresponding
phase diagram is presented in Fig. \ref{fig:phase_diagram}.
In CMR
compounds, the bandwidth $W_0 \gg \hbar \omega_0$, and therefore the summation in Eq. (\ref{eq:F_du}) always
includes two and only two terms. The minimum at small distortions,
$\delta u^{(1)} \simeq \frac{\lambda^2 \hbar \omega_0 }{k c_0 W_0 \langle \cos\frac{\theta}{2} \rangle }$,
which we will refer to as the ``Migdal minimum", corresponds to extended electron eigenstates with
renormalized effective mass { with the} energy
\begin{eqnarray}
E_e \simeq - x \left(1 - x\right) \left[ \frac{c_0}{2} W_\theta -  
\frac{\hbar \omega_0 \lambda^2}{2 k c_0 W_\theta} \right]
\label{eq:E_e}
\end{eqnarray}

The other minimum, which we will refer to as the "Holstein minimum", occurs 
at $u^{(2)} \simeq \lambda / M \omega_0^2$,
and corresponds to small polarons with the energy
\begin{eqnarray}
E_p \simeq - x \left(1 - x\right) \left[ \frac{c_0}{2} 
W_\theta \exp\left( - \frac{\lambda^2}{k \omega_0 \hbar} \right) -
\frac{\lambda^2}{2 k} \right]
\label{eq:E_p}
\end{eqnarray}

Substituting (\ref{eq:E_e}) and (\ref{eq:E_p}) into Eq. (\ref{eq:F_du}), we obtain:
\begin{eqnarray}
F\left[P_\vartheta\right] & = & - T \log\left[ \exp\left( - \frac{E_e}{T}\right)  +
\gamma \exp\left( - \frac{E_p}{T}\right) \right] \nonumber \\
& + & T \int d\vartheta \sin\vartheta P_\vartheta\left(\vartheta\right)
\log P_\vartheta\left(\vartheta\right)
\label{eq:F_m}
\end{eqnarray}
where $\gamma \simeq \sqrt{ 1 + \frac{2 c_0 x \left(1 - x\right)}{1 + 4 x \left(1 - x\right)} \frac{ W_\theta }{\hbar \omega_0}}$

\begin{figure}[t]
\begin{center}
\leavevmode
\epsfxsize = 7. truecm
\epsfbox{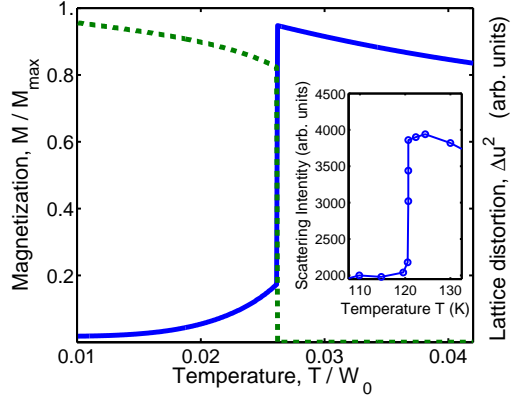}
\end{center}
\caption{
The magnetization $M/M_{\rm sat}$ (dashed blue line)
and the lattice distortion $\Delta u$ in units of  $\lambda/k$ (green line) as
functions of temperature $T$ (in units of $W_0$), for the carrier concentration $x = 0.3$.
The inset shows the experimental temperature dependence of diffuse scattering intensity
due to lattice distortion from Ref. \protect\onlinecite{Shimomura99}.
\protect\label{fig:MUT}
}
\end{figure}

The equations (\ref{eq:F_m}), (\ref{eq:E_e}) and (\ref{eq:E_p}) define a variational
problem for the spin distribution function $P_\vartheta$, which can be
solved by a straightforward minimization
of the free energy.
The corresponding procedure is similar to the one used for the
standard DE model and described
in detail in Ref. \onlinecite{NV00}.
We generally find a phase transition between low lattice distortion ferromagnetic phase
at low temperatures and high lattice distortion paramagnetic phase at hight temperatures.
For a strong electron-lattice coupling, $k W_0 \lesssim \lambda^2$ ,
the transition is
of the first kind, and is accompanied by the virtual collapse of the lattice distortion
$\delta u$, as shown in Fig. \ref{fig:MUT}. Note the remarkable
similarity of the predicted behavior
to the observations in recent experiments \cite{Shimomura99,Vasiliu99},
as illustrated by the inset.

\begin{figure}[t]
\begin{center}
\leavevmode
\epsfxsize = 8. truecm
\epsfbox{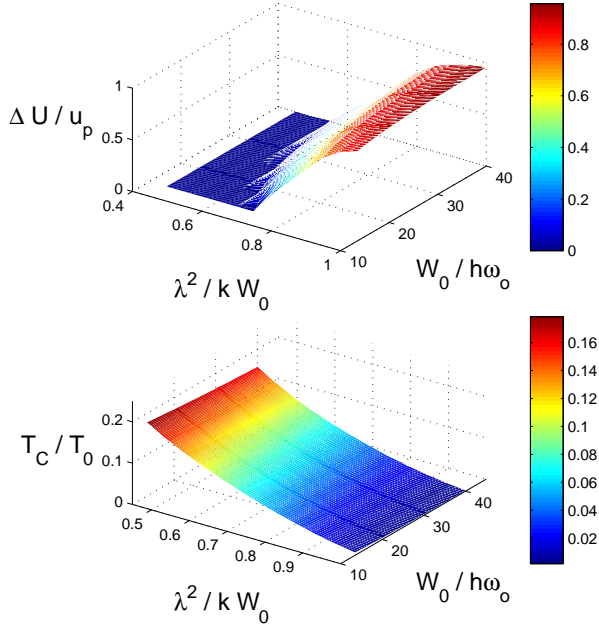}
\end{center}
\caption{
The change of lattice distortion at the transition in units of $u_p \equiv \lambda/k$ (top panel) and the 
critical temperature in units of 
$T_0 \equiv \frac{c_0}{2} x (1 - x) W_0$ (bottom panel), 
as functions of
$\lambda^2 / (k W_0)$ and $W_0/\hbar\omega_0$, in false-color representation. 
The ``deep blue sea" in the top panel represents the
parameters for the second-order transition when $\Delta u \equiv 0$. The carrier concentration
$x = 0.3$.
\protect\label{fig:3D}
}
\end{figure}

In the other limit, when the electron-lattice coupling is small ( i.e.
$ k W\hbar \omega_0 \gg  \lambda^2$), our model corresponds to the standard double-exchange,
and the transition is of the second kind. In fig. \ref{fig:3D} (a) we plot the change of the lattice distortion
$\delta u \equiv u_2 - u_1$ at the phase transition, $\Delta u|_c$,
 as a function of the model parameters, $ \lambda^2/ k W_0$ and
$W_0 / \hbar\omega_0$, in false-color representation. There, the ``deep blue sea" for small electron-lattice
coupling corresponds to the second-order phase transition, where $\Delta u|_c \equiv 0$ by definition.

The phase transition changes from the first order kind to the second order when the ratio of the polaron energy
$E_p \equiv \lambda^2/k$ to the electron bandwidth $W_0$ is near unity. This is in a remarkable agreement with
experimental data, as these quantities in lanthanum manganites are believed to be of the same order, {\it and}
both first\cite{Shimomura99,Vasiliu99} and second-order phase transitions\cite{Tokura} are indeed observed in experiments depending on the composition
(which is likely to affect the electron-lattice coupling and the bandwidth).

An increase of the electron-lattice coupling leads to suppression of the
metallic phase, which manifests itself in smaller values of the critical temperature.
This behavior is illustrated in the bottom panel of Fig. \ref{fig:3D} , where we present the false-color representation
of the dependence $T_c$ as a function of both parameters of the model,
$\lambda^2/ k W_0$ and $W_0 / \hbar\omega_0$.

\begin{figure}[t]
\begin{center}
\leavevmode
\epsfxsize = 7. truecm
\epsfbox{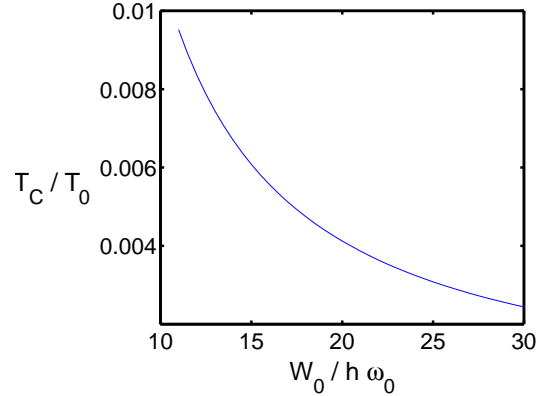}
\end{center}
\caption{
The critical temperature as a function of $W_0 / \hbar \omega_0 \sim \sqrt{M}$ 
(for a fixed value of $\lambda^2 / (k W_0) = 1.$).
\protect\label{fig:tc_kp}
}
\end{figure}

An important feature of the proposed theory is the strong dependence of the critical temperature of the ferromagnetic transition
on the effective ionic mass $M$. This behavior is illustrated in Fig. \ref{fig:tc_kp} where we plot the critical temperature (in units of the
``bare" bandwidth $W_0$) vs. $\frac{k}{\hbar \omega_0} \lambda^2 / k^2 \sim \sqrt{M}$. As follows from Fig. \ref{fig:tc_kp}, a 10\% change
in the oxygen ion mass which corresponds to the replacement of ${}^{16}O$ by ${}^{18}O$, could account for a 
variation of critical
temperature by $~20\%$, which explains the recently observed giant isotope effect in lanthanum manganites.\cite{Zhao96}

Note the remarkable ``anti-correlation" of the change of the lattice distortion at the transition
with the critical temperature, as seen from Fig. \ref{fig:3D}.
This finding is again in agreement with the experimental data\cite{Tokura,Shimomura99,Vasiliu99} :
the transition of the second order
kind is generally observed in the compounds with relatively high values of $T_c$, 
while a second-order transition
is found in materials with a smaller value of critical 
temperature.

The dependence of $P_u(u_1,u_2)$ on $\Delta u$ in our ansatz for
the distribution function  (\ref{eq:p_product}) introduces the essential
correlation of lattice distortion 
with electronic occupation. If this correlation is ignored by approximating
$P_u(u_1,u_2)$  in a mean-field fashion by $p(u_1)p(u_2)$, the important 
results of our theory presented in Figs. \ref{fig:MUT}-\ref{fig:tc_kp} which
include the collapse of the lattice disorder at first-order ferromagnetic 
transition and strong isotope effect in the critical temperature, 
cannot be obtained.

We conclude with a discussion of the effect of intrinsic disorder.
The intrinsic disorder, which can be represented in the model
Hamiltonian (\ref{eq:H}) by a term of the type $\sum_i V_i c_i^\dagger c_i$,
favors less ordered paramagnetic phase and therefore
leads to a decrease of the critical temperature. This effect was considered
in detail in Ref. \onlinecite{NV00}. More important is the effect
of the intrinsic disorder on the transport properties of the CMR compounds.
In high quality samples with relatively weak disorder, the
``Migdal minimum" describes the extended states and is responsible for
the metallic behavior below the transition temperature.  

If the electron-phonon coupling is substantial, the situation
is dramatically different above the transition where the dominant
 contribution is given by the ``Holstein minimum". There, the effective bandwidth
is exponentially suppressed due to lattice distortions
\begin{eqnarray}
W_{H} & \propto & W_0 \exp\left( - \frac{\lambda^2 /k} {\hbar \omega_0} v^2 \right)
\label{eq:W_H}
\end{eqnarray}
where $\lambda^2/k \sim W_0 \gg \hbar \omega_0 $ (in manganites $W_0/\hbar \omega_0 \sim 50$)
and $v \equiv k \ \delta u / \lambda$ is the dimensionless lattice distortion which in
the paramagnetic phase is of the order unity (see Fig. \ref{fig:MUT}).
Therefore $W_{H} \ll W_0$, and small intrinsic disorder in the CMR compounds
will lead to the localization of the Holstein polarons.
On the other side of the ferromagnetic transition, small intrinsic disorder alone, 
because of the vanishing spin as well as polaron disorder, is insufficient to 
localize the electronic states. So metallic conduction occurs below the 
ferromagnetic transition, while an activated form of resistivity is obtained
above the transition, in agreement with experimental data. 

In contrast, for low electron-phonon coupling constants,
as evidenced for example by high (second-order) transition temperatures,
in pure samples the resistivity is expected to be "metallic" on both sides of the
transition. Experimental results are consistent with these observations.

To summarize, in this paper a systematic method has been found
 to isolate and calculate the
principal effects of electron-lattice interactions in strongly correlated
materials of the double-exchange family. By including effects 
of the correlation of lattice vibrations with local electron occupation, 
we have shown that
for the same reason that the itinerant state sweeps away spin-disorder, it also sweeps 
away lattice disorder. This has been accomplished by a variational form for the
couple electron-lattice that extrapolates from the Holstein form to the Migdal form
as the self-consistently determined bandwidth changes from a value small compared to
polaron binding to a value large compared to it. The results of the calculations
 explain many of the peculiar features of the experimental results of CMR compounds.

We gratefully acknowledge the hospitality of the Aspen Center for Physics where a part of this work was done.
E.N. would like to thank A.~Tkachenko for helpful discussions.

\end{document}